\documentclass[prl,aps,amsmath,amssymb,superscriptaddress,twocolumn]{revtex4}
\pdfoutput=1

\usepackage{graphicx}

\newcommand{\nc}{\newcommand}
\nc{\beq}{\begin{equation}}
\nc{\eeq}{\end{equation}}
\nc{\beqa}{\begin{eqnarray}}
\nc{\eeqa}{\end{eqnarray}}

\def\gsim{\mathrel{\rlap{\lower4pt\hbox{\hskip1pt$\sim$}}
    \raise1pt\hbox{$>$}}}       %greater than or approx. symbol

\begin{document}

\title{Entanglement and Fast Quantum Thermalization in Heavy Ion Collisions}
\author{Chiu Man Ho} \email{cmho@msu.edu}
%\affiliation{}
\author{Stephen~D.~H.~Hsu} \email{hsu@msu.edu}
\affiliation{Department of Physics and Astronomy \\ Michigan State University \\  }

\begin{abstract}
Let $A$ be subsystem of a larger system $A  \cup B$, and $\psi$ be a typical state from the subspace of the Hilbert space ${\cal H}_{AB}$ satisfying an energy constraint. Then $\rho_A(\psi)= {\rm Tr}_B \vert \psi \rangle \langle \psi \vert$ is nearly thermal. We discuss how this observation is related to fast thermalization of the central region ($\approx A$) in heavy ion collisions, where $B$ represents other degrees of freedom (soft modes, hard jets, collinear particles) outside of $A$. Entanglement between the modes in $A$ and $B$ plays a central role: the entanglement entropy $S_A$ increases rapidly in the collision. In gauge-gravity duality, $S_A$ is related to the area of extremal surfaces in the bulk, which can be studied using gravitational duals.
\end{abstract}
%\pacs{}
\maketitle
\date{today}

%\bigskip
%\noindent {\bf Introduction}
%\bigskip

%\section{Introduction}

The usual particle scattering description of thermalization in heavy ion collisions (HIC) considers individual scattering events for each degree of freedom (i.e., parton). These scattering events are implicitly assumed to be {\it incoherent} -- largely independent of each other. In the standard kinetic theory picture, multiple scatterings of each parton, one after another, are required to bring their distribution to the high entropy thermal configuration. By the usual reasoning, a timescale larger than 1 fm is required in order for these multiple, successive scatterings to occur.

However, nucleus by nucleus scattering in HIC is a case of coherent scattering of two objects, each with many internal degrees of freedom. We can, in principle, describe the evolution of the wave functional $\psi$ of the entire two nucleus system from asymptotic separation at early times to some specific post-collision moment in time, such as $\tau \approx 1$ fm after the initial overlap of the two nuclei. The wave functional then describes a {\it superposition} of each possible set of interactions or scatterings of individual partons in the system. The bulk or macroscopic properties of the remnant post-collision system are determined by an average over this complex superposition. Only a fraction of the total number of degrees of freedom are in the central region of the collision. The density matrix describing its properties is obtained through a trace over the remainder of the complex state.

Recent results in the foundations of quantum statistical mechanics suggest that the system can thermalize through the spread of entanglement (i.e., superposition) much faster than expected from the usual (incoherent) kinetic theory. That is, the system can evolve from a very atypical state (two heavy nuclei, each with large boost) to a more typical one (entangled superposition state) over a short timescale.

Below, we discuss fast thermalization in HIC (i.e., on timescales of $\tau \approx 1$ fm or less) in relation to the complex nature of the entangled superposition state, and the resulting growth of entanglement entropy. To be precise, what requires explanation is not full thermalization of the central region $A$, but rather the weaker condition of isotropization so that a hydrodynamic description becomes approximately valid. At minimum, the stress tensor of the matter in $A$ must assume a diagonal form in its rest frame. For an overview, see \cite{iso}.

What is new in this paper: we relate fast thermalization in HIC to recent results in the foundations of quantum statistical mechanics (the properties of {\it typical} states). We conjecture that entanglement between different branches of the HIC wavefunction is the primary driver of fast thermalization -- it is an intrinsically quantum effect, not a semi-classical one. This conjecture is supported by gauge-gravity duality: the Ryu-Takayanagi formula relates the area of extremal surfaces to entanglement entropy in HIC, implying a large increase in the latter during the collision.

\bigskip
\noindent {\bf Quantum thermalization and properties of typical pure states}
\bigskip

%\section{Quantum thermalization and properties of typical pure states}

{\it Typical} pure states in quantum mechanics are states which dominate the Hilbert measure. The ergodic theorem proved by von Neumann \cite{vN1,vN2} implies that under Schrodinger evolution most systems spend almost all their time in typical states. Typical states are maximally entangled, and the approach to equilibrium can be thought of in terms of the spread of entanglement.

\begin{figure}
\includegraphics[width=8cm]{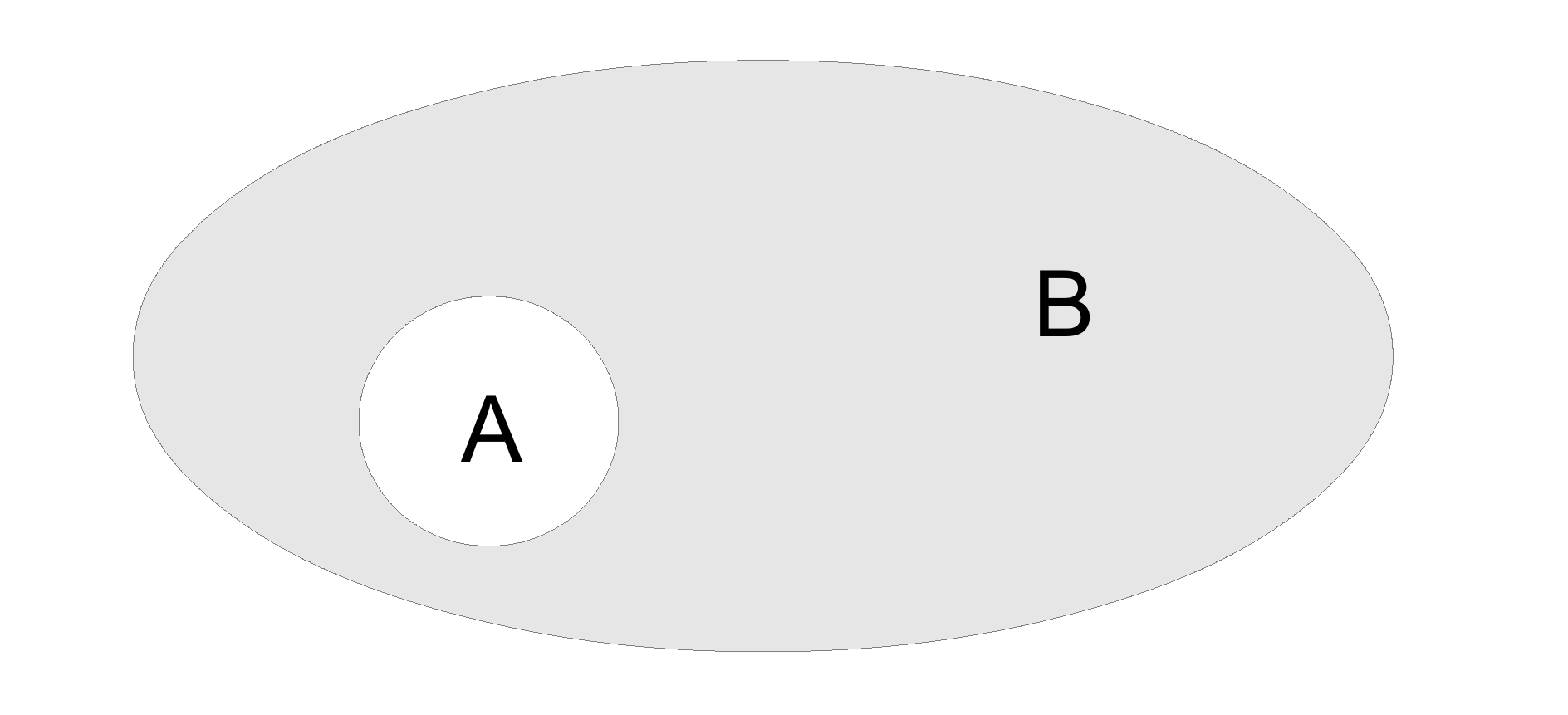}
\caption{The entire system is in a pure state $\psi$ subject to a constraint on total energy. Tracing over the shaded region B yields a density matrix $\rho_A$. For typical $\psi$ (which dominate the set of possible states), $\rho_A$ is nearly thermal.}
\label{AB}
\end{figure}

Consider a large system subject to a linear constraint $R$ (for example, that it be in a superposition of energy eigenstates with the energy eigenvalues all being near some $E_\ast$), which reduces its Hilbert space from ${\cal H}$ to a subspace ${\cal H}_R$. Divide the system into a subsystem $A$, to be measured, and the remaining degrees of freedom which constitute an environment $B$ (Fig. \ref{AB}), so ${\cal H} = {\cal H}_A \otimes {\cal H}_B$ and
 \begin{equation}
 \rho_A \equiv \rho_A(\psi)= {\rm Tr}_B \vert \psi \rangle \langle \psi \vert\,,
\end{equation}
is the density matrix which governs measurements on $A$ for a given pure state $\psi$ of the whole system. The entanglement entropy $S_A$ is $S (\rho_A) = - \,{\rm Tr} \, \rho_A \ln \rho_A\,$.

It can be shown \cite{Winters1,intuition} (see also \cite{Gemmer,Goldstein}), using the concentration of measure on hyperspheres \cite{HDG} (Levy's Lemma), that for almost all $\psi \in {\cal H}_R$,
\begin{equation}
\rho_A(\psi) \approx {\rm Tr}_B\left(\rho_*\right) \equiv \Omega_A\,,
\end{equation}
where $\rho_* = { \bf 1}_R / d_R$ is the equiprobable maximally mixed state on the restricted Hilbert space ${\cal H}_R$ (${ \bf 1}_R$ is the identity projection on ${\cal H}_R$ and $d_R$ the dimensionality of ${\cal H}_R$). $\Omega_A = {\rm Tr}_B \left(\rho_*\right)$ is the corresponding canonical state of the subsystem $A$. The result holds as long as $d_B \gg d_A$, where $d_{A}$ and $d_{B}$ are the dimensionalities of  the ${\cal H}_A$ and ${\cal H}_B$ Hilbert spaces. (Recall that these dimensionalities grow exponentially with the number of degrees of freedom. The Hilbert space of an $n$ qubit system is $2^n$--dimensional.) In the case of an energy constraint $R$, $\Omega_A$ describes a perfectly thermalized subsystem with temperature determined by the total energy of the system (i.e., a micro canonical ensemble).

To state the theorem in \cite{Winters1} more precisely, the (measurement-theoretic) notion of the \emph{trace-norm} is required, which can be used to characterize the distance between two mixed states $\rho_A$ and $\Omega_A$:
\begin{equation}
\Vert\rho_A-\Omega_A\Vert_1\equiv{\rm Tr}\sqrt{\left(\rho_A-\Omega_A\right)^2}\,.
\end{equation}
This quantifies how easily the two states can be distinguished by measurements, according to the identity
\begin{equation}
\label{tracesup}
\Vert\rho_A-\Omega_A\Vert_1 = {\rm sup}_{\Vert O\Vert\leq1}\,{\rm Tr}\left(\rho_AO-\Omega_AO\right)\,,
\end{equation}
where the supremum runs over all observables $O$ with operator norm $\Vert O\Vert \leq$ 1. The trace on the right-hand side of (\ref{tracesup}) is the difference of the observable averages $\langle O\rangle$ evaluated on the two states $\rho_A$ and $\Omega_A$, and therefore specifies the experimental accuracy necessary to distinguish these states in measurements of $O$.

The theorem then states that (for $\epsilon > 0$)
\begin{align}
\label{w1}
& {\rm Prob} \left[
\Vert\rho_A\left(\psi\right)-{\rm Tr}_B\left(\rho_*\right)\Vert_1 ~\geq~ \epsilon+ d_A  d_R^{\, -1/2}
\right]  \nonumber \\
&~~~~~~~~~~~~~~~~~ <~  2\exp ( -\epsilon^2d_R/18\pi^3 )~.
\end{align}
In words: let $\psi$ be chosen randomly (according to the Haar measure on the Hilbert space) out of the space of allowed states ${\cal H}_R$;
the probability that a measurement on the subsystem $A$ \emph{only}, with measurement accuracy given in (\ref{w1}), will be able to tell the pure state $\psi$ (of the entire system) apart from the maximally mixed state $\rho_*$ is exponentially small in $d_R$, the dimension of the space ${\cal H}_R$ of allowed states. Conversely, for almost all pure states $\psi$ any small subsystem $A$ will be found to be extremely close to perfectly thermalized (assuming the constraint $R$ on the whole system was an energy constraint).

As mentioned, the overwhelming dominance of typical states $\psi$ is due to the geometry of high-dimensional Hilbert space and the resulting concentration of measure. It is a consequence of kinematics only -- no assumptions have been made about the dynamics. Almost any dynamics -- i.e., choice of Hamiltonian and resulting unitary evolution of $\psi$ -- leads to the system spending nearly all of its time in typical states for which the density matrix describing any small subsystem $A$ is nearly thermal \cite{vN1,vN2,Winters2}. Typical states $\psi$ are maximally entangled (i.e., $S_A$ is nearly maximal), and the approach to equilibrium can be thought of in terms of the spread of entanglement, as opposed to the more familiar non-equilibrium kinetic equations, which describe incoherent scattering.

Explicit demonstrations of fast quantum thermalization have been obtained for broad classes initial states, on timescales of order the inverse temperature (i.e., average energy per mode) \cite{fast1,fast2}. Indeed the conceptual challenge is to understand the conditions which lead to the more familiar slow (semi-classical) thermalization.

Since generic pure states tend to evolve into typical states, any mixture of pure states is likely to evolve into a mixture of typical states. Hence, our analysis does not require any specific assumptions about whether the system is in a pure or mixed state. If it is in a mixture, we simply have (classical) probabilities of finding the system in one of two or more typical pure states. For simplicity, we can assume the system as a whole is in a pure state.

\bigskip
\noindent {\bf Application to Heavy Ion Collisions}
\bigskip

%\section{Application to Heavy Ion Collisions}

In HIC we let $A$ represent mostly soft particles in the central region, and let $B$ represent all other modes (Fig. 2). In highly central collisions at RHIC, the nucleons comprising each heavy nucleus pass through each other, losing energy due to interactions. Some of this energy is deposited in the central region. The rapidity of the original nucleons drops from about 6 to 5 \cite{Westphal}. Therefore, at most about 40 percent of the total energy in the two nucleus system ends up in the central region, with not all of it registered in detectors. In most collisions, which are peripheral (lower centrality), much less than 40 percent of the total energy is deposited in the central region. So, the dimensionalities $d_{A,B}$ of the relevant Hilbert spaces describing $A$ and $B$ obey the inequality $d_A \ll d_B$.

Thus, the conditions are appropriate for the application of the results discussed above, IF the post collision state $\psi$ at time $\tau \approx 1$ fm after the initial overlap of the two nuclei is sufficiently typical among states of the same energy. Obviously the state is not fully typical -- most of the energy is still in highly boosted nucleons largely collinear with the beamline. However, the complex pure state $\psi$ describing {\it all possible combinatorial scattering trajectories} is clearly highly entangled, and tracing over most of its degrees of freedom generates a large amount of entropy. $\rho_A$ is therefore a high entropy state even if it is not fully thermalized. It seems plausible that $\rho_A$ might be described in macroscopic terms using some kind of hydrodynamic approximation \cite{hydro1,hydro2,hydro3}, even at times as early as $\tau \approx 1$ fm.

Indeed, the quantum thermalization contemplated here generates much more entropy in $\rho_A$ at early times than could arise from the usual picture of incoherent scatterings of individual partons. In the usual picture, one must explain how degrees of freedom with typical energy 200 to 300 MeV (given by the initial conditions used in hydrodynamical simulations \cite{hydro1,hydro2,hydro3}) are able to assume a nearly thermal or at least highly entropic distribution. The timescale $\tau \approx 1$ fm allows at most a single scattering with momentum transfer less than $~ 200$ MeV for each particle, so there is not enough time for fine adjustment of energies or momenta of soft modes.

On the other hand, superposition -- quantum parallelism -- can adjust the probability distribution over soft energies and momenta in the density matrix $\rho_A$. That is, since all possible scatterings occur in the superposition state, the final probability distribution takes into account much finer grained effects.

To better understand the superposition state, consider a parton $i$ in one of the heavy nuclei, with momentum $p_i$. This parton will pass through the opposing nucleus, and (in a given scattering amplitude) might (or might not) scatter against any of a large number of opposing partons. The first scattering of $i$ leads to a range of possibilities for its new momentum $p_i^{(1)}$, and after each subsequent scattering there is another new momentum $p_i^{(n)}$. So a sequence of values
\begin{equation}
\{ p_i^{(1)}, p_i^{(2)}, \cdots p_i^{(n)} \}\,,
\end{equation}
is associated with each parton's trajectory. Inelastic scattering creates many additional particles of lower energy, and each of these new particles has its own trajectory. {\it All possible scattering histories of all partons} appear in the superposition state $\psi$. When most of the degrees of freedom in $\psi$ are traced over, the resulting $\rho_A$ has very high entropy.

Clearly, the amount of entanglement entropy generated increases with the total energy of the collision. At energies below some threshold, $\rho_A$ presumably does not become approximately isotropic or hydrodynamic. Similarly, the strength of interactions plays an important role. In the limit of weak coupling we would not have significant particle creation, nor multiple scatterings per particle. Region $A$ would not be highly populated with soft modes, whose specific state (e.g., energy-momentum distribution) is correlated to that of region $B$. In the earlier discussion of ``typical'' states the coupling does not play a role because the results {\it assume} that the state $\psi$ of the system is typical -- how and whether it {\it becomes} typical depends on the dynamics or strength of interactions.

For earlier work (with a somewhat different perspective) on entanglement entropy and thermalization in strong interactions, see \cite{Elze1,Elze2,Muller}.

\begin{figure}[t!]
\label{HIC}
\includegraphics[width=8cm,height=6cm]{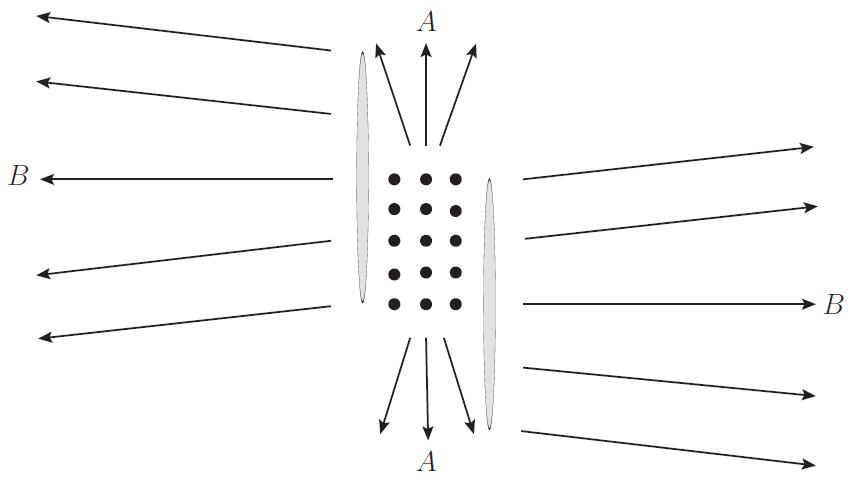}
\caption{The collision at $\approx 1$ fm after crossing of the two Lorentz-contracted nuclei. Hard and collinear modes lie in part B of the Hilbert space. The soft central modes lie in part A. Tracing over all or most of the degrees of freedom in B leads to a high entropy density matrix describing A.}
\end{figure}

\bigskip
%\newpage
\bigskip
\noindent {\bf Conclusion}
\bigskip

%\section{Conclusion}

Fast thermalization in HIC might be a real-world example of a novel and intrinsically quantum mechanical phenomenon. Rapid growth of entanglement in the complex superposition of all possible scattering amplitudes leads to a large entropy $S (\rho_A) = - \,{\rm Tr} \, \rho_A \ln \rho_A\,$ even if thermalization is incomplete. This observation might explain the applicability of hydrodynamical models (isotropization) even at early times such as $\tau \approx 1$ fm.

It would be interesting to explore these ideas further in AdS models, where fast thermalization and colliding gravitational shocks have been studied \cite{Yaffe1,Yaffe2}. The Ryu-Takayanagi (RT) formula relates the entanglement entropy of a gauge state $A$ to the area of the bulk extremal surface $\gamma_A$ which terminates on the boundary of $A$ \cite{AdS1,AdS2}:
\begin{equation}
S_A = { {\rm Area} ( \gamma_A ) \over 4 }\,.
\end{equation}
In bulk collisions which correspond to the HIC (e.g., collisions of gravitational waves), a horizon (black hole) is typically formed, which causes an increase in the area of the extremal surface. 

Thus, the gauge-gravity duality {\it implies} that the entanglement entropy of the region A increases significantly in HIC. That is, the thermalization (or isotropization; i.e., approach to equilibrium) is {\it related to entanglement}. See \cite{AdS1,AdS2,bh1,bh2,pure1,pure2} for more discussion, including examples of condensed matter systems which are explicit realizations of the quantum thermalization phenomenon. In \cite{pure1,pure2} the distinction between von Neumann entropy (which remains zero in the time evolution of a pure state) and entanglement entropy $S_A$ (which is sensitive to horizon formation in the bulk, and changes in time) is discussed in detail in the context of solvable boundary models such as free fermions.

One can consider our paper to be simply an examination of how this entanglement thermalization looks from the gauge/QCD side in a heavy ion collision.

Finally, we noted that fast thermalization on timescales of order the inverse temperature seems to be generic for broad classes of quantum states and Hamiltonians \cite{fast1,fast2}. If, as is sometimes conjectured, all QFTs have gravitational duals, this result might be understood in terms of horizon formation: the timescale for gravitational collapse is of order the light crossing time of the resulting black hole, which is roughly the inverse temperature.

%\newpage
\bigskip
\bigskip

\emph{Acknowledgements---}  The authors acknowledge support from the Office of the Vice-President for Research and Graduate Studies at Michigan State University.

%\newpage

%%%%%%%%%%%%%%%%%%%%%%%%%%%%%%%%%%%%%%%%%%%%%%%%%%%%%%%%%%%%%%%%%
%%%
%%%                     BIBLIOGRAPHY
%%%
%%%%%%%%%%%%%%%%%%%%%%%%%%%%%%%%%%%%%%%%%%%%%%%%%%%%%%%%%%%%%%%%%

\bigskip

%\newpage
%\vskip .75 in

\end{document}